\documentclass[aps,prl,twocolumn,showpacs,superscriptaddress,groupedaddress]{revtex4}  
\usepackage{graphicx}
\usepackage{dcolumn}   
\usepackage{bm}        
\usepackage{amsmath,amssymb,mathrsfs}
\usepackage{xfrac}
\usepackage{float}
\usepackage{braket}

\newcommand{\abs}[1]{\ensuremath{\left| #1 \right|}}

\begin{document}

\title{Amplification of Angular Rotations Using Weak Measurements}

\author{Omar S. Maga\~{n}a-Loaiza}
\email{omar.maganaloaiza@rochester.edu}

\affiliation{The Institute of Optics, University of Rochester, Rochester, New York 14627, USA}

\author{Mohammad Mirhosseini}
\email{mirhosse@optics.rochester.edu}
\affiliation{The Institute of Optics, University of Rochester, Rochester, New York 14627, USA}

\author{Brandon Rodenburg}
\email{brandon.rodenburg@gmail.com}
\affiliation{The Institute of Optics, University of Rochester, Rochester, New York 14627, USA}

\author{Robert~W.~Boyd}
\affiliation{The Institute of Optics, University of Rochester, Rochester, New York 14627, USA}
\affiliation{Department of Physics, University of Ottawa, Ottawa ON K1N 6N5, Canada}

\begin{abstract}
    We present a weak measurement protocol that permits a sensitive estimation
    of angular rotations based on the concept of weak-value amplification. The
    shift in the state of a pointer, in both angular position and the conjugate
    orbital angular momentum bases, is used to estimate angular rotations.
    This is done by an amplification of both the real and imaginary parts of
    the weak-value of a polarization operator that has been coupled to the
    pointer, which is a spatial mode, via a spin-orbit coupling. Our experiment
    demonstrates the first realization of weak-value amplification in the
    azimuthal degree of freedom. We have achieved effective amplification
    factors as large as 100, providing a sensitivity that is on par with more
    complicated methods that employ quantum states of light or extremely large
    values of orbital angular momentum.
\end{abstract}

\maketitle

In 1988 Aharonov \emph{et al.\!}~\cite{Aharonov:1988fk} introduced a general
form of quantum measurement, known as a weak measurement. In weak measurements,
information is gained by weakly probing the system, while approximately
preserving its initial state.  The uncertainty in each measurement is large due
to the weak perturbative nature of the information extraction; however, this is
generally overcome by averaging over a large number of identically prepared
states. The process of post-selecting the prepared system makes weak
measurements interesting. Under certain conditions the outcome, which is called
a weak value (WV), is not an eigenvalue of the measurement operator. In fact,
WVs can even exceed the eigenvalue range of a typical strong or projective
measurement and in general are complex. These features have allowed a wide
range of applicability in classical and quantum contexts. For example, they
have resulted in the measurement via amplification of small
transverse~\cite{Dixon:2009eu, Hosten:2008ih} and
longitudinal~\cite{Brunner:2010a, Xu:2013fu, Strubi:2013hb} shifts, the direct
measurement of the quantum wave function~\cite{Lundeen:2011is, Salvail:2013bo,
Malik:2014bf}, the development of tomographic
techniques~\cite{Kobayashi:2013tj}, the amplification of optical
nonlinearities~\cite{Feizpour:2011bs}, and the clarification of controversial
debates in quantum physics~\cite{Kocsis:2011jg, Aharonov:2002wt}.

Recently, there has been a strong impetus to employ weak-value amplification
(WVA) as an effective tool in metrology ~\cite{Brunner:2010a, Xu:2013fu,
SalazarSerrano:2014uo, Hofmann:2011gg}. A WVA protocol involves the preparation
of an ensemble of particles with two independent degrees of freedom (DoF).
These two DoFs are then coupled by means of a weak perturbation and
post-selected to collapse one of the DoF, typically called the probe.  Due to
the coupling existing between the probe and the other DoF, called the pointer,
the post-selection induces a shift in the linear position of the pointer which
is proportional to the weakly induced perturbation and the WV. This has allowed
the use of WVA to estimate small quantities with sensitivities comparable to
quantum-enhanced metrology~\cite{Brunner:2010a, Xu:2013fu,
SalazarSerrano:2014uo, Hofmann:2011gg, Giovannetti:2011jk}, due to the fact
that the use of quantum protocols does not guarantee sensitivities beyond the
standard quantum limit, which is the limit for classical
protocols~\cite{ThomasPeter:2011dg, Shin:42140}.

Besides the extensive work on the estimation of longitudinal
displacements~\cite{Brunner:2010a, Xu:2013fu, Strubi:2013hb,
SalazarSerrano:2014uo, ThomasPeter:2011dg, Shin:42140, Higgins:2007tc}, high
sensitivity measurement of angular displacements has been another topic of
interest. Historically, inquiries regarding relativistic dynamics stimulated
interest on the azimuthal DoF \cite{THOMAS:1926gv}. A remarkable example is the
Sagnac effect. Atomic versions of the Sagnac interferometer have led to
sensitive gyroscopes that permit a precise measurement of
rotations~\cite{Gustavson:1997vo, Stockton:2011hy}. In addition, the use of
light endowed with orbital angular momentum (OAM) has motivated interest in new
forms of rotations. As identified by Allen \emph{et al.\!}~\cite{Allen:1992vk},
an optical beam with azimuthal phase dependence of the form $e^{i\ell\phi}$
carries OAM, where $\phi$ is the azimuthal angle and $\ell$ is the OAM value.
These beams have been used for rotational control of microscopic systems
\cite{Padgett:2011ju}, and exploration of effects such as the rotational
Doppler shift \cite{Courtial:1998vj} which has been recently used in techniques
for detecting spinning objects~\cite{Lavery:2013iz, RosalesGuzman:2013fh}.
Recent efforts to increase the sensitivity in the measurement of angular
rotations involve the generation of large values of
OAM~\cite{Ambrosio:2013fsa}, quantum entanglement of high OAM
values~\cite{Fickler:2012vr}, or the use of N00N states in the OAM
bases~\cite{Jha:2011it}. These protocols require complicated schemes to
generate and measure photons in such exotic states. However, the concepts
behind them constitute valuable resources not only for optical metrology,
remote sensing, biological imaging or navigation systems~\cite{Lavery:2013iz,
RosalesGuzman:2013fh, UribePatarroyo:2013fj}, but also for the understanding of
light-matter interactions~\cite{Gorodetski:2012cg, Luo:2008ub,
FrankeArnold:2011ec}.

\begin{figure*}[!ht]
    \centering    
    \includegraphics[scale=0.0869]{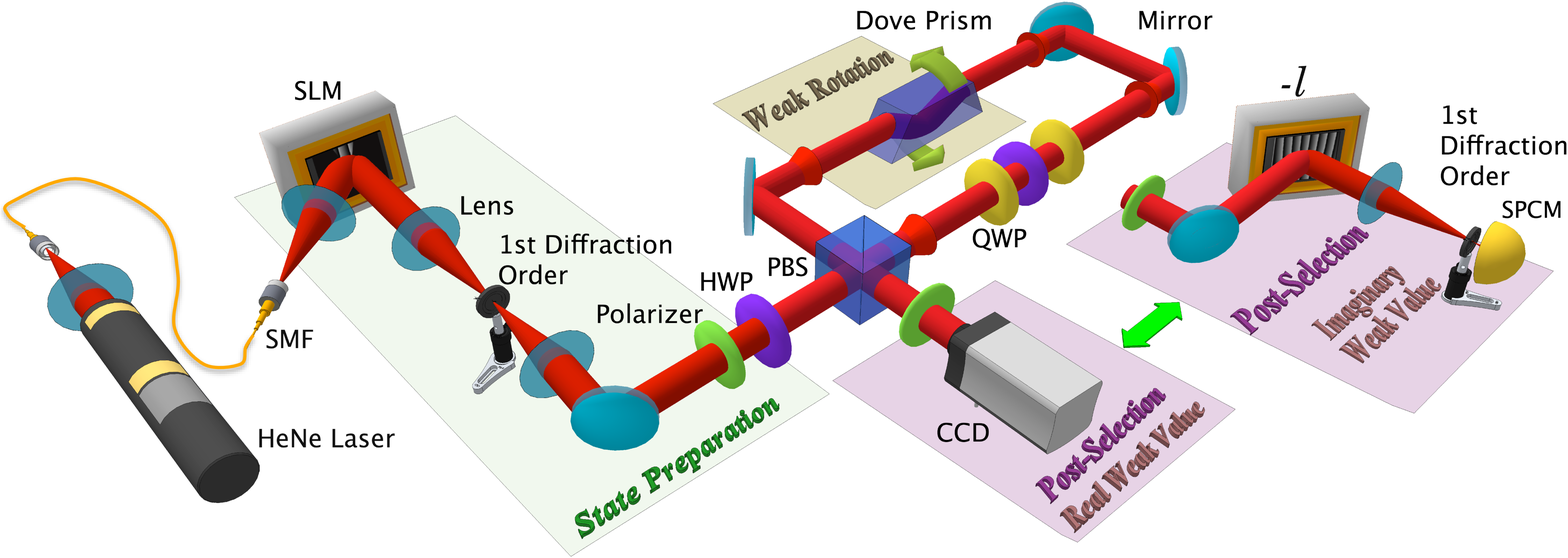}
    \caption{Experimental Setup. A light beam from HeNe laser working at 632.8
        nm is coupled into a single-mode fiber (SMF) and the output is then collimated. The
        beam is sent to a phase-only spatial light modulator (SLM) and then to
        a 4f optical system containing a spatial filter in the Fourier plane.
        A polarization state is prepared by means of a polarizer and a
        half-wave plate (HWP). A dove prism (DP), a HWP and two quarter-wave
        plates (QWP) are placed inside the Sagnac interferometer. The DP
        induces a small rotation between the counter-propagating beams; this is
        the weak perturbation. The QWPs together with the HWP induce a
        geometric phase between the H and V polarized beams. After
        post-selection, measurements of angular rotations and OAM spectra are
        performed to access the real or imaginary part of the weak-value.}
    \label{fig:ExpSetup} 
\end{figure*}

In this work, we describe WVA in the azimuthal DoF and the processes that give
rise to these effects. The first observation of these kinds of WVs suggests
interesting physics from the fundamental and applied perspective. For instance,
the spin-orbit coupling in our experiment gives rise to an interesting optical
effect in which the perturbation of polarization induces a shift in the angular
position and OAM spectrum of the pointer. We show that the real and the
imaginary part of the WV for the polarization operator can be accessed by
measuring the angular position and its conjugate variable of OAM, respectively.
Using this new form of WVs based on spin-orbit coupling, we propose a scheme
for the measurement of small rotations. We demonstrate an amplification in the
measurement of angular rotations that is as large as 100. The simplicity of our
scheme, namely lack of need for exotic quantum state of lights or extremely
large values of OAM, makes this technique potentially attractive for
applications in optical metrology, remote sensing and optical manipulation of
microscopic systems.

\begin{figure*}[!ht]
    \centering
    \includegraphics[width=0.726\textwidth]{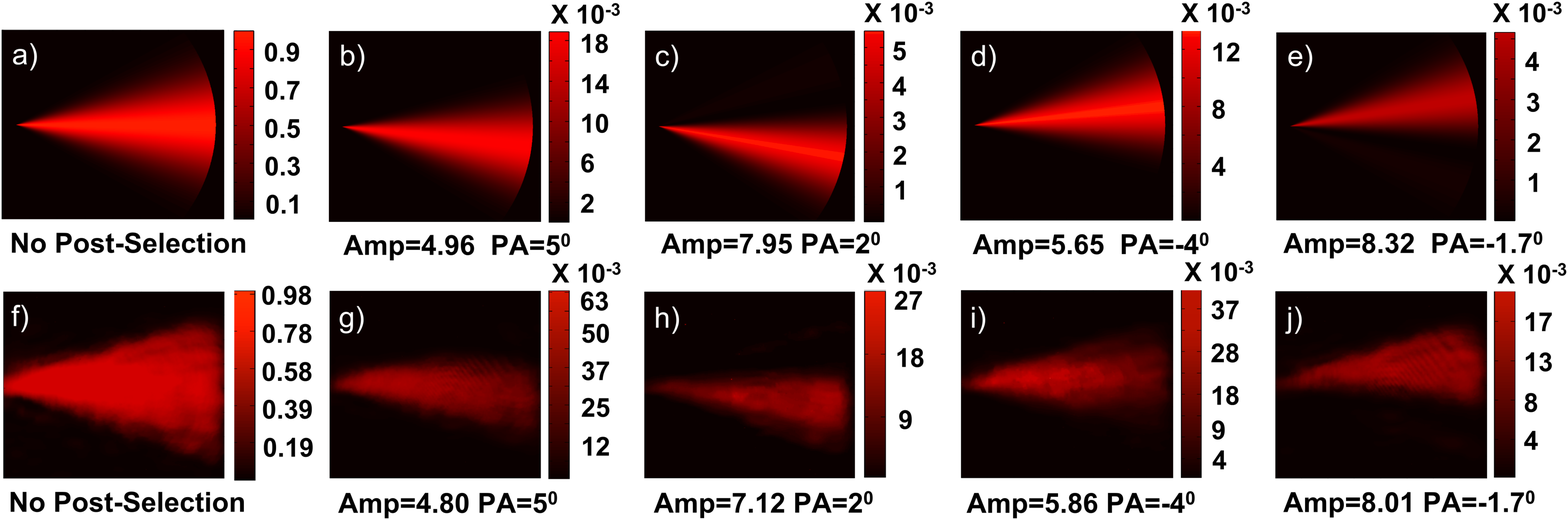}
    \caption{Amplification of angular displacements using real weak values.  a)
        -- e) show simulations of our scheme for $\Delta\phi=1.2^\circ$,
         different post-selection angles (PA) and amplification factors (Amp).
     f) -- j) show experimental evidence of our protocol under the same
 conditions.}
    \label{fig:RealWV}
\end{figure*}
Consider the experimental setup depicted in Fig.~\ref{fig:ExpSetup}. This
scheme comprises three parts: state preparation, a weak perturbation, and
post-selection. The state preparation involves the generation of a light beam
with diagonal polarization and a well-defined spatial profile. We select the
initial polarization state using a polarizer and a half-wave plate (HWP); this
state will serve as a probe and can be described by the polarization qubit
$\ket{\Psi_{pr}} = \frac{1}{\sqrt{2}}(\ket{H}+\ket{V})$. The preparation of the
spatial mode or pointer consists of the generation of an angular mode (ANG)
$f(\phi)\propto \exp{(-\phi^2/2{\eta_{\phi}}^2)}$, which is a Gaussian-apodized
angular slit of width $\eta_{\phi}$. This is shaped by impressing amplitude and
phase information onto the beam by means of modulation of the blaze parameters
on a spatial light modulator (SLM), used together with a 4f optical system
containing a spatial filter in the Fourier plane~\cite{Davis:1999ku}.  The
advantages that pointer states carrying OAM provide over Gaussian pointer
states have been studied \cite{Puentes:2012jr}. The beam is injected into a
Sagnac interferometer, where the horizontally and vertically polarized
components of the beam circulate in opposite directions.  The dove prism (DP)
is rotated by a small angle $\Delta\phi/4$ with respect to the plane of the
interferometer, which causes the two counter-propagating beams to be rotated by
an amount of $\pm\Delta\phi/2$ in opposite directions. This setup enables a
coupling between the polarization, marked by the two counter propagating beams,
and the transverse azimuthal DoF.  In the next step we use two quarter-wave
plates (QWP) and a HWP to induce a geometric phase between the two circulating
beams in the interferometer, permitting the existence of complex WVs
\cite{Sup_Inf_I}. Finally, the post-selection is carried out by setting the
angle of a polarizer almost orthogonal with respect to the angle of the
polarizer used in the pre-selection. At this stage, a full characterization of
the complex wavefunction in the transverse angular basis and the conjugate
basis of OAM reveals information about the real and the imaginary part of the
WV, respectively. 

The interaction  in our experiment can be described by the spin-orbit
interaction Hamiltonian $\hat{H}_{SO} = \mu\hat{\sigma}\hat{\ell}_z$ and a
Hamiltonian that describes the action of the wave plates
$\hat{H}_g=\delta\hat{\sigma}$, where $\hat{\sigma}$ is the Pauli operator
defined by $\hat{\sigma} \equiv \ket{H}\bra{H} - \ket{V}\bra{V}$,
$\frac{\Delta\phi}{2}=\mu\Delta t$,
$(\frac{\theta_H}{2}-\frac{\pi}{2})=\delta\Delta t$ and $\frac{\theta_H}{2}$ is
the induced geometric phase. Our state at the input of the interferometer has
the following form $\ket{\Psi_i} = \ket{\Psi_{pr}}\ket{f\left(\phi\right)}$.
The interaction which occurs in the DP couples the two DoFs as follows:
\begin{equation}
\begin{split}
& \ket{\Psi_f}
    = e^{-i\frac{\Delta\phi}{2}\hat{\sigma}\hat{\ell}_z}e^{-i\hat{\sigma}(\frac{\theta_H}{2}-\frac{\pi}{2})}\Ket{\Psi_i} \\ 
&= \frac{1}{\sqrt{2}}\left( e^{-i\frac{\theta}{2}}\Ket{H}\Ket{f\left(\phi-\Delta\phi/2\right)}
                     +e^{i\frac{\theta}{2}}\Ket{V}\Ket{f\left(\phi+\Delta\phi/2\right)}\right),
\end{split}
\end{equation} 
where \(\hat{\ell}_z\) act as the generator of rotations and is proportional to
the angular momentum operator projected along the optical axis
$\hat{L}_z=\hbar\hat{\ell}_z$, and $\theta$ equals $\theta_{H}-\pi$. As
demonstrated by Eq. 1, the weak coupling creates entanglement between probe and
pointer. It should be noted that since the probe and the pointer are different
DoFs of a single beam rather than separate systems or particles, then this is
an example of classical entanglement and thus can be described
classically~\cite{Qian:2011vu, Kagalwala:2012jo}. Because of this, most
traditional weak measurement experiments, such as those described in
Refs.~\cite{Dixon:2009eu, Brunner:2010a, Lundeen:2011is, Salvail:2013bo,
Malik:2014bf, Kobayashi:2013tj, Xu:2013fu, Puentes:2012jr}, are classically
explainable. This also demonstrates what is required to perform a non-classical
weak measurement experiment. We have chosen to use the mature language of weak
measurement theory, since it provides a simpler description and the results
readily apply to a wider range of phenomena including non-classical systems.

The post-selection is performed by projecting the perturbed state into
$\ket{\Phi_{ps}} = \sin\left(\frac{\gamma}{2} - \frac{\pi}{4}\right)\ket{H}
+\cos\left(\frac{\gamma}{2} - \frac{\pi}{4}\right)\ket{V}$, where $\gamma$ is
controlled by the polarizer. The post-selection collapses the polarization
state of the probe and causes a shift in the angular position and the OAM
spectrum of the pointer that can be described as
\begin{equation}
    \ket{\Psi_p}=\ket{\Phi_{ps}}\Braket{\Phi_{ps}|\Psi_f}
    \approx \ket{\Phi_{ps}}\Ket{f\left(\phi- \sigma_{w}\Delta\phi/2\right)}.
    \label{eqn:Psi_p}
\end{equation}
Here, $\sigma_{w}$ is the complex WV given by
\begin{eqnarray}
   \sigma_w &\equiv&  \frac{\Braket{\Phi_{ps}|\hat{\sigma}|\Psi_{fpr}}}{\Braket{\Phi_{ps}|\Psi_{fpr}}} 
\end{eqnarray}
$\ket{\Psi_{fpr}}$ is defined as
$\frac{1}{\sqrt{2}}\left(e^{-i\frac{\theta}{2}}\ket{H}+e^{i\frac{\theta}{2}}\ket{V}\right)$.
If the induced phase $\theta$ and polarizer selection angle $\gamma/2$ are
small, the WV can be approximated as \cite{Sup_Inf_II}
\begin{equation}
    \sigma_w \approx -\frac{2\gamma}{{\gamma}^2+{\theta}^2}+i\frac{2\theta}{{\gamma}^2+{\theta}^2}.
    \label{eqn:ApproxWeakValue}
\end{equation}

The post-selected state described in Eq.~\ref{eqn:Psi_p} shows a change in
angle as $\phi\rightarrow\phi-\sigma_w\Delta\phi/ 2$. If $\sigma_w$ is real,
which will be the case for $\theta=0$, then this leads to the rotation of the
pointer by the amount $\sigma_w$. However if $\sigma_w$ is complex then
\begin{equation}
\begin{split}
   \label{eqn:ImagWeakValue}
   & f(\phi-\sigma_w\Delta\phi/2) =
    e^{\left(-(\phi-\sigma_w\Delta\phi/2)^2/2\eta_\phi^2\right)} \\
   &\qquad\propto e^{\left(-(\phi-\Re{(\sigma_w)}\Delta\phi/2)^2/2\eta_\phi^2\right)}
     e^{\left(i\phi\Im{(\sigma_w)}\Delta\phi/2\eta_\phi^2\right)} \\
    &\qquad = e^{\left(-(\phi-\Delta\Braket{\phi})^2/2\eta_\phi^2\right)}
        e^{\left(i\phi\Delta\Braket{\ell}\right)}, 
\end{split}        
\end{equation}
where $\Delta\Braket{\phi} = \Re{(\sigma_w)}\Delta\phi/2$ sets the amount of
the pointer's rotation. In addition, the pointer experiences a shift in its OAM
spectrum that equals $\Delta\Braket{\ell} =
\Im{(\sigma_w)}\Delta\phi/2\eta_\phi^2$. We have used the angular
representation of the spatial mode of the photons, and utilized the Fourier
relation between the conjugate pairs of azimuthal angle and angular momentum.
Alternatively, the same results can be derived by using the commutation
relation between angular position and OAM operators, which is given by
$[\hat{\phi},\hat{L}_z]=i\hbar(1-2\pi P(\phi))$ where $P(\phi)$ represents the
angular probability at the boundary of the angle
range~\cite{FrankeArnold:2004kc}. The shift in the OAM spectrum can be
understood as a form of interaction between spin angular momentum (SAM) and
OAM. This interesting optical effect is a consequence of the
polarization-sensitive nature of the interactions in the interferometer, and
should not be confused with the standard spin-orbit coupling in the vector
beams where both the SAM and OAM are directed along the same axis
\cite{Nagali:2009jz}.

\begin{figure*}[!ht]
    \centering    
    \includegraphics[scale=0.77]{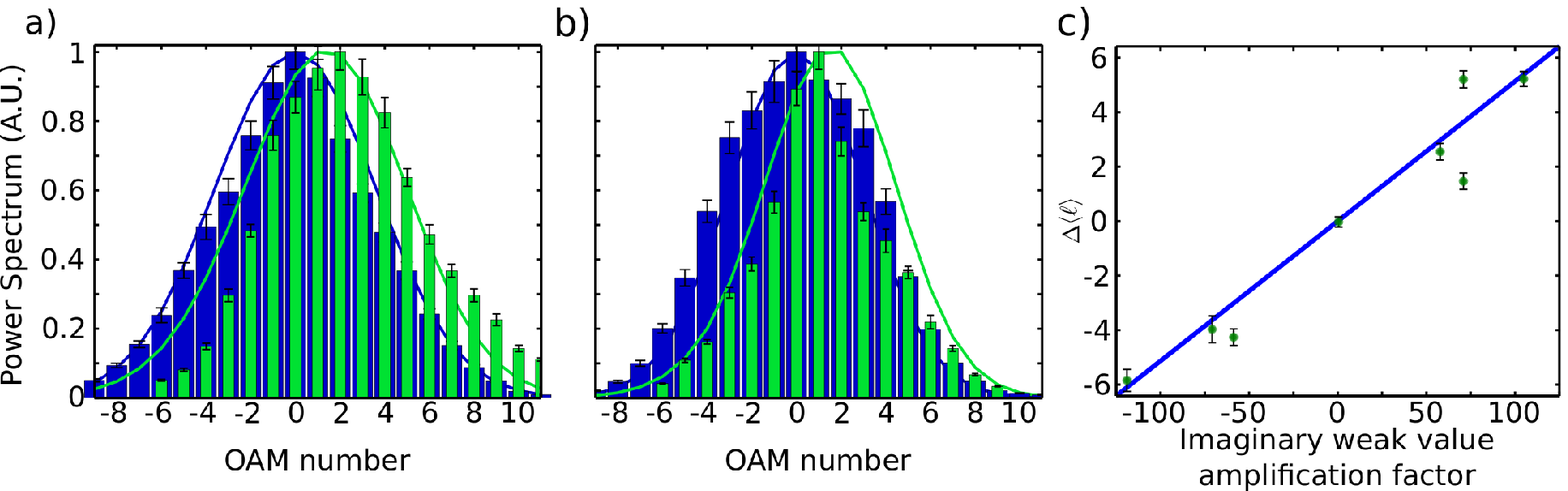}
    \caption{Measured OAM power spectra of $\ket{\Psi_p}$ without
    post-selection (blue) and with post-selection (green) demonstrating the
    shift in $\braket{\ell}$ due to $\Im{(\sigma_w)}$ for a)
    $\eta_\phi=11.4^0$, $\gamma/2=6^0$ and b) $\eta_\phi=13.7^0$ and
    $\gamma/2=5^0$. The angle $\theta/2$ equals $5^0$ for all the cases.
    Histograms represent measured data, while lines represent theoretically
    predicted shifts. c) OAM centroid shift $\Delta\Braket{\ell}$ for various
    measured OAM power spectra plotted against the imaginary WV amplification
    factor, $\Im{(\sigma_w)}/2\eta_\phi^2$. Dots represent data, while the line
    is the theoretical linear curve predicted by Eq.~\ref{eqn:ImagWeakValue}.}
    \label{fig:ImagWV}
\end{figure*}

In the experiment we use a 3 mW He-Ne laser (632.8 nm) which is coupled to a
single-mode fiber (SMF) and then expanded to a spot size of 1.8 cm. The central
part of the beam homogeneously illuminates the display of the SLM that has an
active area of \( 9.3 \times 7 \text{mm}^2 \). Due to the reflectance of the
SLM and the efficiency of the encoded diffractive grating on it, the power
drops to 470 nW once an ANG mode of width $\eta_{\phi}=13.7^\circ$ is
generated.  The DP in the Sagnac interferometer is rotated by $0.3^\circ$, this
angle is determined by measuring a relative rotation of $1.2^\circ$ between two
identical ANG modes propagating in the opposite directions. The induced
displacement $\Delta\phi$, is chosen to be much smaller than the width of the
ANG mode, in order to guarantee the conditions for the weak perturbation. The
post-selection polarizer is set to an angle $\gamma/2$, with respect to the
polarization state of the pre-selected state. For this part, we have set
$\theta$ to zero.

Since our interest is in the amplification of the weak-value, the angle
$\gamma/2$ is set to a small number. The post-selection polarizer forces the
two ANG modes to coherently interfere, producing another ANG mode which is
rotated due to the azimuthal Gaussian intensity distribution impressed in the
ANG~\cite{Ritchie:1991gh}. Such rotation is proportional to the angular
displacement $\Delta\phi$ and the real part of the WV $\Re{(\sigma_w)}$. Since
the WV can take values larger than one, this scheme allows the amplification of
small rotations. However, as $\Re{(\sigma_w)}$ is increased more photons are
lost as shown for different post-selection angles (PA) in
Fig.~\ref{fig:RealWV}a--e. In order to detect this effect, a CCD camera is
placed after the polarizer. This is equivalent to measuring the expected value
of the angular position in the state $\ket{\Psi_p}$. As shown in
Fig.~\ref{fig:RealWV}f--j, the measured power is in the range of 10--30 nW,
however these images were taken using long exposure times. As can be seen in
Fig.~\ref{fig:RealWV}, an aggressive post-selection leads to a larger rotation.
The amplification factor (Amp) is defined as the ratio between the angular
position of the post-selected mode $\Delta \langle \phi \rangle$ and
$\Delta\phi$. This is equal to $\Re{(\sigma_w)}/2$.  Both $\Delta \langle \phi
\rangle$ and $\Delta\phi$ were determined by using centroid measurements. The
amplification limit is given by the extinction ratio of the polarizer and the
magnitude of the weak perturbation or the angle of post-selection. Larger
amplifications can be measured if the width of ANG is increased and the
post-selection angle is decreased. 

The imaginary part of the WV can be determined by analyzing the shift of the
OAM spectrum of the ANG. We have chosen the rotation angle of the DP to be
approximately $0.4^\circ$ and we have tried different angular widths for the
input state. In order to allow $\Im{(\sigma_w)}$ to be nonzero, the phase
$\theta$ must also be nonzero. This is done by inducing a geometric phase
between the polarization states $\Ket{H}$ and $\Ket{V}$. This phase is created
using three rotatable wave plates as shown in Fig.~\ref{fig:ExpSetup}.  The
angle of the QWPs is set to $\pi/4$ and the HWP is rotated by a small angle
\cite{Sup_Inf_I}. We have set the HWP to an angle such that $\theta/2 =
5^\circ$ and tried several different post-selection angles for the polarizer.
Measurement of the OAM spectrum associated with a beam can be done using a wide
variety of techniques~\cite{Mair:2001fd, Leach:2002wy, Berkhout:2010cb,
Mirhosseini:2013em}. We measured the OAM using a series of projective
measurements for various values of $\ell$.  Using a similar procedure as was
used for generating the angular slits, a hologram was impressed onto a SLM and
then a Fourier transforming lens and a spatial filtering from a SMF couples
photons to an APD which allows measurement at single photon levels
\cite{Sup_Inf_III}. 

We summed the counts during a 0.2 second window and averaged it for 30
measurements for each projection over different OAM modes.  This procedure was
repeated for each mode and the reconstructed spectra are shown in Fig.~3a--b.
The error bars represent the standard deviation over the ensemble of 30
measurements. The spectrum is broader for angular modes with narrower widths
due to uncertainty relation between angular position and OAM
\cite{FrankeArnold:2004kc}. As predicted by Eq.~\ref{eqn:ImagWeakValue}, and
shown in Fig.~\ref{fig:ImagWV}, the larger amplifications are obtained for
angular modes with narrow widths. However, such narrow ANG modes have
physically smaller cross sections and hence carry proportionally less power.
Each OAM power spectra was fitted using a weighted least-squares minimization
to a shifted Gaussian function.  The mean values are plotting in
Fig.~\ref{fig:ImagWV}c along with error bars representing the $3\sigma$
confidence interval. By exploiting the measurement process we have amplified
small rotations by a magnitude of 100 without using high OAM nor entanglement. 

Recently, there has been research casting doubt on the sensitivity of
measurements based on WVA \cite{Ferrie:2014un}. However, it has been shown that
in the presence of technical noise WVs, and more specifically imaginary WVs,
outperform traditional measurements \cite{Jordan:2014jv}. Therefore, our
experiment can be potentially useful for sensitive measurement of small
rotation in real world scenarios. A quantitative analysis of the sensitivity of
our scheme can be done by using the Fisher information metric. However,  such
quantitative comparison is outside the scope of the current work and will be
the subject of a future study.

We have made the first step towards the study of WVA in the azimuthal DoF. This
has been approached by describing the mechanisms that lead to a shift in the
angular position and OAM of an optical beam. The OAM spectrum is shifted as a
consequence of the breakup in the polarization symmetry, realized by a
differential geometric phase.  Furthermore, we have implemented the first
realization of WVA in the angular position and OAM bases. The results presented
here provide a proof-of-principle demonstration of the scope of WVA in this
DOF. We believe that our protocol opens the possibility for new schemes in
optical metrology.  In addition, our approach shows an alternative fashion to
study the exchange between SAM and OAM in optical systems.

We acknowledge discussions with J. Vornehm, G. Viza, J. Martinez-Rincon, J. C.
Howell and E. Karimi. This work was supported by DARPA, CERC and CONACyT.

\widetext
\begin{center}
\textbf{\large Supplementary information}
\end{center}
\setcounter{equation}{0}
\setcounter{figure}{0}
\setcounter{table}{0}
\setcounter{page}{1}
\makeatletter
\renewcommand{\theequation}{S\arabic{equation}}
\renewcommand{\thefigure}{S\arabic{figure}}
\renewcommand{\bibnumfmt}[1]{[S#1]}
\renewcommand{\citenumfont}[1]{S#1}

\begin{quote}
    The main purpose of this document is to provide more detailed information about
    the design of our experimental setup and recast its functionality in terms of
    the ``weak measurement'' formalism. In addition, we describe the scheme
    employed to measure the OAM spectrum of a beam of light.
\end{quote}

\section*{I. Sagnac interferometer}
We use a Sagnac interferometer composed of a polarized beam splitter (PBS), a
Dove prism (DP) and a series of rotatable plates. In this first section, we will
describe the role of the DP and how we use the rotatable plates to introduce
geometric phases. 

As mentioned in the article, a spatial mode $\ket{f(\phi)}$ is diagonally
polarized and injected into the input port of the interferometer, and the
polarization information is described by the state $\ket{\Psi_{pr}}$. Therefore
we can describe the initial state as $\ket{f(\phi)}\ket{\Psi_{pr}}$. The beam
is split into two polarization components that circulate in opposite directions
within the interferometer. The role of the DP is to rotate the beam. When the
DP is rotated by an angle of $\Delta\phi/4$ about its optical axis, the
transmitted beam suffers a rotation of $\pm\Delta\phi/2$, where the sign is
determined by the propagation direction of the beam. In our experiment, the DP
couples the polarization degree of freedom (DoF) to the spatial DoF of the
beam. The action of the DP inside the polarized Sagnac interferometer is
described as follows:
\begin{equation}
    \begin{array}{cc}
        \ket{\Psi_{pr}}\ket{f(\phi)}\rightarrow[Prism]{Dove}
        &\frac{1}{\sqrt{2}} (\bold{R}(-{\phi}/2)\ket{f(\phi)}\ket{H}+\bold{R}({\phi}/2)\ket{f(\phi)}\ket{V})\\
        &=\frac{1}{\sqrt{2}} \ket{f(\phi-\Delta\phi/2)}\ket{H}+\ket{f(\phi+\Delta\phi/2)}\ket{V}),
    \end{array} 
\end{equation}
where the operator $\bold{R}(\phi/2)$ is given by
$e^{i\hat{\ell}\Delta\phi/2}$. $\hat{\ell}$ is the generator of rotations
and is proportional to the operator representing angular momentum along the
optical axis.

The role of the wave plates is to induce a geometric phase, which we will
describe using the Jones matrix formalism. Here the polarization states are
defined as:
\begin{equation}
    \ket{H}=
    \begin{bmatrix}
        1\\
        0
    \end{bmatrix}\text{ and }
    \ket{V}=
    \begin{bmatrix}
        0\\
        1
    \end{bmatrix}.
\end{equation}
The action of the quarter-wave plate $\bold{QWP}$ and half-wave plate $\bold{HWP}$ are described by the following matrices,
\begin{equation}
\begin{split}
\begin{array}{l}
    \bold {QWP}=\\ \\ \qquad
    \left[ {\begin{array}{cc}
            \mathrm{e}^{i\phi_{x1}}{\cos}^2(\theta_Q)+\mathrm{e}^{i\phi_{y1}}{\sin}^2(\theta_Q) 
            &( \mathrm{e}^{i\phi_{x1}}-\mathrm{e}^{i\phi_{y1}})\cos(\theta_Q)\sin(\theta_Q) \\
            ( \mathrm{e}^{i\phi_{x1}}-\mathrm{e}^{i\phi_{y1}})\cos(\theta_Q)\sin(\theta_Q)
            &\mathrm{e}^{i\phi_{x1}}{\sin}^2(\theta_Q)+\mathrm{e}^{i\phi_{y1}}{\cos}^2(\theta_Q) \\
    \end{array} } \right],
\end{array}
\end{split}
\label{QWP}
\end{equation}
and
\begin{equation}
    \bold{HWP}=
    \left[ {\begin{array}{cc}
        \cos(2\theta_H) & \sin(2\theta_H)\\
        \sin(2\theta_H) & -\cos(2\theta_H)\\
    \end{array} } \right].
\label{HWP}
\end{equation}
$\theta_{Q}$ represents the orientation of the fast axis of the QWP,
with respect to the x-axis, and $\theta_{H}$ represents the orientation of the
HWP. The value $\phi_y-\phi_x$ determines the induced retardation phase between
the two components of the electric field. For a QWP this number is equal to
$\pi/2$.

The configuration used to induce the geometric phase consists of a HWP
sandwiched between two QWPs.  The angle $\theta_{Q}$ was set to $\pi/4$ whereas
the angle of the HWP, $\theta_{H}$, was set to $\theta_{H}/4$. Thus the
Eqs.~\ref{QWP}--\ref{HWP} become
\begin{equation}
    \bold {QWP_{\pm}}=
    \left[ {\begin{array}{cc}
        \frac{1}{2}+\frac{i}{2}      &\pm(\frac{1}{2}-\frac{i}{2})\\
        \pm(\frac{1}{2}-\frac{i}{2}) &\frac{1}{2}+\frac{i}{2}\\
    \end{array} } \right],
\end{equation}
and
\begin{equation}
    \bold{HWP_{\pm}}=
    \left[ {\begin{array}{cc}
        \cos(\theta_H/2)        &\pm\sin(\theta_H/2)\\
        \pm \sin(\theta_H/2)    & -\cos(\theta_H/2)\\
    \end{array} } \right].
\end{equation}
Different orientation angles have to be considered for each of the
counter-propagating beams. Here we use positive and negative values for the
horizontally and negative polarized beams respectively. The transformation
suffered by each beam is thus given by
\begin{equation}\label{xx}
\begin{split}
    \ket{H^{g}}
    &=\bold{QWP_+} \cdot \bold{HWP_+} \cdot \bold{QWP_+}\ket{H}\\
    &=e^{-i(\theta_{H}/2-\pi/2)}\ket{H},
\end{split} 
\end{equation} 
and
\begin{equation}\label{xx}
\begin{split}
\ket{V^{g}}&=\bold{QWP_-} \cdot \bold{HWP_-} \cdot \bold{QWP_-}\ket{V}\\
&=e^{i(\theta_{H}/2-\pi/2)}\ket{V}.
\end{split} 
\end{equation} 
As can be seen, the net effect is the acquisition of a phase given by
$\pm(\theta_{H}/2-\pi/2)$, which is the geometric phase.

\section*{II. The action of sagnac interferometer in terms of the weak
measurement formalism}

In this section we describe our weak measurement protocol. The action of the
interferometer is described by the following interaction Hamiltonian:
\begin{equation}\label{xx}
    \hat{H}_T=\hat{H}_g+\hat{H}_{SO}.
\end{equation} 
The Hamiltonian $\hat{H}_g$ describes the role of the three wave plates. It is
given by $\delta\hat\sigma$. The spin-orbit interaction caused by the DP is
described by $\hat{H}_{SO}$, which is given by a Hamiltonian of the form
$\mu\hat\sigma\hat\ell$, where $\hat\sigma$ is the Pauli operator defined as
$\ket{H}\bra{H}-\ket{V}\bra{V}$, $(\theta_{H}/2-\pi/2)=\delta\Delta t$ and
$\Delta\phi/2=\mu\Delta t$. Given this, the evolution of the initial state
$\ket{\Psi_i}=\ket{\Psi_{pr}}\ket{f(\phi)}$ to a final state $\ket{\Psi_f}$ 
is given as
\begin{equation}\label{xx}
\begin{split}
    \ket{\Psi_f}&=e^{-i\hat{H}_T\Delta t}\ket{\Psi_{pr}}\ket{f(\phi)}\\
    &=e^{-i\hat{H}_g\Delta t}\Bigg[\frac{1}{\sqrt{2}}(\ket{H}+\ket{V})\ket{f(\phi)}\\
    &\ -i\frac{\Delta\phi}{2\sqrt{2}}{(\ket{H}\bra{H}-\ket{V}\bra{V})}(\ket{H}+\ket{V})\hat\ell\ket{f(\phi)}+\ldots\Bigg],\\
\end{split}
\end{equation} 
This  expression can be rewritten as 
\begin{equation}\label{xx}
    \begin{array}{l}
        \ket{\Psi_f}=\\ \\ \qquad
        \frac{e^{-i\hat{H}_g\Delta t}}{\sqrt{2}}\left[\ket{H}(1-i\frac{\Delta\phi}{2}\hat\ell+...)\ket{f(\phi)}
        +\ket{V}(1+i\frac{\Delta\phi}{2}\hat\ell+...)\ket{f(\phi)}\right].
    \end{array}
\end{equation} 
The expression in parenthesis is a translation operator in the azimuthal degree
of freedom which leads to the state
\begin{equation}\label{xx}
    \ket{\Psi_f}=
    \frac{e^{-i\hat{H}_g \Delta t}}{\sqrt{2}}[\ket{H}\ket{f(\phi-\Delta\phi/2)}
    +\ket{V}\ket{f(\phi+\Delta\phi/2)}].
\end{equation} 
The action of $\hat{H}_{g}$ leads to the state
\begin{equation}\label{xx}
\begin{split}
    \ket{\Psi_f}&=
    \frac{1}{\sqrt{2}}[e^{-i(\theta_H/2-\pi/2)}\ket{H}\ket{f(\phi-\Delta\phi/2)}\\
    &\qquad +e^{i(\theta_H/2-\pi/2)}\ket{V}\ket{f(\phi+\Delta\phi/2)}].
\end{split}
\end{equation} 

The state above describes our experiment just before post-selection by the
polarizer is performed. $\Ket{\Psi_f}$ is the state of the photons emerging
from the output port of the polarized beam splitter.  The post-selection
process is described by the projection operator
$\ket{\Phi_{ps}}\bra{\Phi_{ps}}$ which gives the post-selected state
\begin{equation}\label{xx}
\begin{split}
\ket{\Psi_p}&=\ket{\Phi_{ps}}\bra{\Phi_{ps}}\ket{\Psi_f}\\
&=[\bra{\Phi_{ps}}e^{-i\hat{H}_g \Delta t}\ket{\Psi_{pr}}\ket{f(\phi)}\\
&\qquad-i\frac{\Delta\phi}{2}\bra{\Phi_{ps}}e^{-i\hat{H}_g \Delta t}\hat\sigma\ket{\Psi_{pr}}\hat\ell\ket{f(\phi)}+...]\ket{\Phi_{ps}}.
\end{split}
\end{equation} 
This expression can be approximated to the first order and then normalized:
\begin{equation}\label{xx}
\ket{\Psi_p}\approx\left(\ket{f(\phi)}-i\frac{\Delta\phi}{2}\frac{\bra{\Phi_{ps}}e^{-i\hat{H}_g \Delta t}\hat\sigma\ket{\Psi_{pr}}}{\bra{\Phi_{ps}}e^{-i\hat{H}_g \Delta t}\ket{\Psi_{pr}}}\hat\ell\ket{f(\phi)}\right)\ket{\Phi_{ps}}
\end{equation} 
Since $\hat{\sigma}$ commutes with $e^{-i\hat{H}_g \Delta t}$
\begin{equation}\label{xx}
\ket{\Psi_p}=\left(\ket{f(\phi)}-i\frac{\Delta\phi}{2}\frac{\bra{\Phi_{ps}}\hat\sigma\ket{\Psi_{fpr}}}{\braket{\Phi_{ps}|\Psi_{fpr}}}\hat\ell\ket{f(\phi)}\right)\ket{\Phi_{ps}},
\end{equation} 
where $\Ket{\Psi_{fpr}}\equiv e^{-i\hat{H}_g \Delta t}\ket{\Psi_{pr}}$.
Defining the weak value of $\hat{\sigma}$ as
\begin{equation}
    \sigma_w\equiv\frac{\bra{\Phi_{ps}}\hat\sigma\ket{\Psi_{fpr}}}{\braket{\Phi_{ps}|\Psi_{fpr}}},
\end{equation}
then the total effect of the post-selection can be written as
\begin{equation}\label{xx}
\ket{\Psi_p}=\ket{f(\phi-\sigma_w\Delta\phi/2)}\ket{\Phi_{ps}}.
\end{equation} 

The weak value of the polarization operator can be determined by using the
following form for the states:
\begin{equation}\label{xx}
\begin{split}
\ket{\Psi_{fpr}}&=\frac{1}{\sqrt{2}}\left(e^{-i(\theta_H/2-\pi/2)}\ket{H} +e^{i(\theta_H/2-\pi/2)}\ket{V}\right)\\
\ket{\Phi_{ps}}&=\sin\left(\gamma/2-\pi/4\right)\ket{H} +\cos\left(\gamma/2-\pi/4\right)\ket{V}.
\end{split}
\end{equation} 
It is worth noting that $\ket{\Phi_{ps}}$ is almost orthogonal with respect to
$\ket{\Psi_{pr}}$. Using the above states the weak value becomes
\begin{equation}\label{xx}
\sigma_w=\frac{\tan\left(\gamma/2-\pi/4\right)e^{-i(\theta_H-\pi)}-1}{\tan\left(\gamma/2-\pi/4\right)e^{-i(\theta_H-\pi)}+1}.
\end{equation} 
For simplicity we can define the angle $\theta_H-\pi$ as $\theta$. If we assume
that $\gamma/2$ and $\theta$ are very small, corresponding to the weak
measurement regime, then this expression becomes
 \begin{equation}\label{xx}
 \begin{split}
\sigma_w&=\frac{e^{-i\theta}\frac{\tan(\gamma/2)-\tan(\pi/4)}{1+\tan(\gamma/2)\tan(\pi/4)}-1}{e^{-i\theta}\frac{\tan(\gamma/2)-\tan(\pi/4)}{1+\tan(\gamma/2)\tan(\pi/4)}+1}\\
&\approx\frac{(1-i\theta)\frac{\gamma-2}{2+\gamma}-1}{(1-i\theta)\frac{\gamma-2}{2+\gamma}+1}\\
&=\frac{\gamma-2-i\theta\gamma+2i\theta-\gamma-2}{\gamma-2-\theta\gamma+2i\theta+\gamma+2}\\
&\approx\frac{-2}{\gamma+i\theta}=-2\frac{\gamma}{\gamma^2+\theta^2}+2\frac{i\theta}{\gamma^2+\theta^2}.
\end{split}
\end{equation} \\

\section*{III. Projective measurements}
In this section we describe a simple form of measuring the OAM power spectrum
of an ensemble of photons. The technique employed is known as projection
measurements. Here a spatial mode (in our case an angular mode $f(\bold{r})$),
which can be written in terms of the modal expansion
$\sum_{\ell}a_{\ell}e^{i\ell\phi}$, is imaged onto a SLM that transforms the
field in the first diffracted order to:
\begin{equation}
f(\bold{r})e^{-i\ell\phi}.
\end{equation} 
A Fourier transforming lens takes the field above to
\begin{equation}
    f_\ell(\boldsymbol{\rho})=\mathscr{F}[{f(\mathbf{r})e^{-i\ell\phi}}],
\end{equation} 
which is spatially filtered using a SMF coupled to an avalanche photodiode
(APD) which allows measurement at single photon levels. The coupling efficiency
into the fiber, $\eta_\ell$ is given by
\begin{equation}
    \eta_\ell\propto\abs{\int f_\ell(\boldsymbol{\rho})e^{-\frac{\rho^2}{2\eta^2}} \, \mathrm{d}^2\boldsymbol{\rho}}^2,
\end{equation} 
where $\eta$ is the Gaussian width of the fiber mode. Assuming features of
$f_\ell$ to be of size scale larger than $\eta$, this filtering function
becomes
\begin{equation}
    \eta_\ell\approx\abs{\int f_\ell(0)e^{-\frac{\rho^2}{2\eta^2}} \, \mathrm{d}^2\boldsymbol{\rho}}^2\propto |f_\ell(0)|^2
    = \abs{\sum_{\ell'}{\int{a_\ell e^{i(\ell'-\ell)\phi}\mathrm{d}^2\bold{r}}}}^2 =
    \abs{a_\ell}^2,
\end{equation} 
permitting us to obtain the OAM power spectrum component $\abs{a_{\ell}}^2$.
This process is repeated for the different modes contained in the spatial mode.
The efficiency of this technique for different spatial modes, such as radial
modes, has been studied in reference~\cite{Qassim}.

\renewcommand\refname{

\end{document}